\documentclass[twocolumn,aps,pre,reprint,superscriptaddress,longbibliography]{revtex4-1}
\usepackage{amsmath,amssymb}\usepackage{graphicx}\usepackage{subfigure}\usepackage{color}
\usepackage{times}\usepackage{tabularx}

\begin{document} 
\title{\bf {Analytic Theory of Finite-size Effects in Supercell Modelings of Charged Interfaces}}
\author{Cong Pan}
\affiliation{State Key Laboratory of Supramolecular Structure and Materials, Jilin University, Changchun, 130012, P. R. China}
\affiliation{College of Chemistry, Jilin University, Changchun, 130012, P. R. China}
\author{Shasha Yi}
\affiliation{State Key Laboratory of Supramolecular Structure and Materials, Jilin University, Changchun, 130012, P. R. China}
\affiliation{College of Chemistry, Jilin University, Changchun, 130012, P. R. China}
\author{Zhonghan Hu} \email{zhonghanhu@jlu.edu.cn}
\affiliation{State Key Laboratory of Supramolecular Structure and Materials, Jilin University, Changchun, 130012, P. R. China}
\affiliation{College of Chemistry, Jilin University, Changchun, 130012, P. R. China}
\date{\today}
\begin{abstract}
The Ewald3D sum with the tinfoil boundary condition (e3dtf) evaluates the electrostatic energy of a finite unit cell inside an infinitely periodic supercell.
Although it has been used as a {\it de facto} standard treatment of electrostatics for simulations of extended polar or charged interfaces, the finite-size effect on
simulated properties has yet to be fully understood.
There is, however, an intuitive way to quantify the average effect arising from the difference between the e3dtf and Coulomb potentials on the response of mobile
charges to contact surfaces with fixed charges and/or to an applied external electric field.
While any charged interface formed by mobile countercharges that compensate the fixed charges slips upon the change of the acting electric field, the distance
between a pair of oppositely charged interfaces is found to be nearly stationary, which allows an analytic finite-size correction to the amount of countercharges.
Application of the theory to solvated electric double layers (insulator/electrolyte interfaces) predicts that the state of complete charge compensation is invariant
with respect to solvent permittivities, which is confirmed by a proper analysis of simulation data in the literature.
\end{abstract}
\maketitle

\section{Introduction}
Reliable all-atom molecular dynamics or Monte-Carlo simulations of condensed phases require a careful treatment of the long-ranged electrostatic interactions among
full or partial atomic charges\cite{Yi_Hu2015}. As the range of the coulomb force is greater than the box length of a typical simulation cell containing $N <
10^7$ charges, truncation methods often produce unacceptable artifacts\cite{Yi_Hu2015,Steinbach_Brooks1994,Feller_Brooks1996,Yonetani2005}. Instead, the routinely
used Ewald3D sum with the tinfoil boundary condition (e3dtf) method\cite{Ewald1921,DeLeeuw_Smith1980,Cisneros_Sagui2014} or its various computationally efficient
implementations\cite{Frenkel_Smit2002} include interactions of the charges with all their periodic images in a supercell. One may express the e3dtf electrostatic
energy of $N$ point charges $(q_j,{\mathbf r}_j)$ with $j=1,2,\cdots,N$ in a simple cubic cell with volume $V=L_xL_yL_z$ as\cite{Hu2014ib,Yi_Hu2017pairwise}
\begin{equation} {\cal U}^{\rm e3dtf} = \sum_{i < j }^N q_i q_j \nu^{\rm e3dtf} ({\mathbf r}_{ij}), \end{equation}
with the pairwise e3dtf potential given by\cite{notepairwise}
\begin{equation}  \nu^{\rm e3dtf}({\mathbf r}) = \frac{\tau^{\rm 3D}}{4\pi\varepsilon_0} + \frac{1}{\varepsilon_0V} \sum_{{\mathbf k}\neq{\mathbf 0}}
\frac{e^{i{\mathbf k}\cdot{\mathbf r}}}{k^2} , \label{eq:pwnu}  \end{equation}
where the constant $\tau^{\rm 3D}$ depends on the geometry of the unit cell\cite{Yi_Hu2017pairwise}. $\varepsilon_0$ is the vacuum permittivity and the reciprocal
vector ${\mathbf k}=2\pi(k_x/L_x,k_y/L_y,k_z/L_z)$ with $k_x$, $k_y$, and $k_z$ integers.
$\nu^{\rm e3dtf}({\mathbf r})$ approaches the free space Coulomb interaction, $\nu_0({\mathbf r})=1/(4\pi\varepsilon_0 r)$ at ${\mathbf r}\to {\mathbf 0}$ but
deviates slowly away from it as ${\mathbf r}$ increases. For any nonuniform charge density, $\nu^{\rm e3dtf}({\mathbf r})$ produces a periodic electric field that
explicitly depends on $L_x$, $L_y$ and $L_z$. Unless the simulation is carried out in an infinitely large unit cell for which the difference between $\nu^{\rm
e3dtf}({\mathbf r})$  and $\nu_0({\mathbf r})$ vanishes, one may expect strong finite-size effects on simulated quantities of nonuniform polar/charged systems, which
however is commonly ignored in the literature\cite{Yi_Hu2015}.

It turns out that the spatial symmetry of a nonuniform system plays a crucial role in determining the effect imposed by the finiteness of the simulation
box\cite{Pan_Hu2017}. For a confined planar (spherical) system, the difference between the e3dtf method and the rigorous treatment---the Ewald2D sum
method\cite{Parry1975, Heyes_Clarke1977,DeLeeuw_Perram1979,Pan_Hu2014,Pan_Hu2015} (the Coulomb interaction)---yields on average the so-called planar (spherical)
infinite boundary term of the Coulomb lattice sum\cite{DeLeeuw_Smith1980,Smith1981,Hu2014ib}. 
For the dielectric response of liquids confined between two planar walls, the mean-field equation accounting for the average effect arising from the planar infinite
boundary term must lead to a finite-size correction to the acting electric field.
In contrast, for spherical systems under a spherical electric field, the corresponding correction arising from the spherical infinite boundary term fortuitously
cancels. 
These mean-field equations verified by examples of confined water systems, however become inapplicable for extended systems where the boundary of the unit cell
bisects a material instead of the vacuum region making the infinite boundary terms ill-defined\cite{Pan_Hu2017}. 

\begin{figure*}[tbhp]
\centering\includegraphics[width=11.4cm]{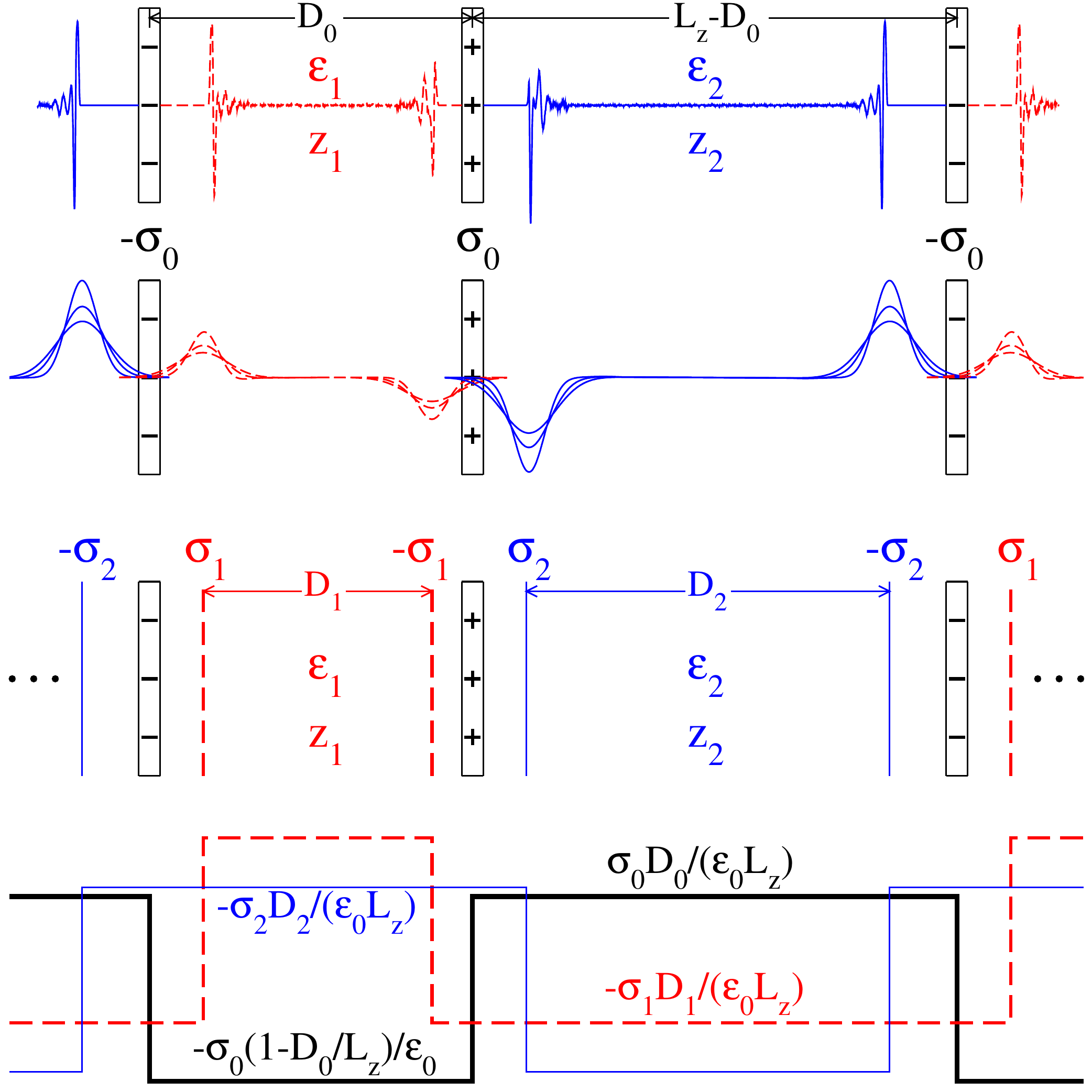} 
\caption{Periodic model of two liquid dielectrics with relative permittivities of $\varepsilon_1$ and $\varepsilon_2$ responding to contact charged walls and an
applied constant electric field. Narrow columns in the upper, middle-upper and middle-lower panels refer to the fixed walls with surface charge densities of
$\pm\sigma_0$ and their periodic images.  The fixed walls are located in the $xy$ plane. The applied (app) external electric field (not shown) is along the $z$
direction with a magnitude of $E_{\rm app}$. $z_1$ and $z_2$ denote $z$ positions of the two bulk regions respectively.
Periodic mobile charge densities of the two liquids, $\rho^q_1(z)$ and $\rho^q_2(z)$ are depicted by red dash and blue thin solid lines respectively in the upper 
panel, along with the corresponding periodic smoothed charge densities, $\rho^{q,\lambda}_1(z)$ and $\rho^{q,\lambda}_2(z)$ with three values of $\lambda$ and the
corresponding periodic interfaces of countercharges in the middle-upper and middle-lower panels respectively.
The distances between a pair of oppositely charged interfaces of countercharges, $D_1$ and $D_2$ are nearly constant with respect to the change of 
the electric field, $E_{\rm app}$ or the fixed surface charge density, $\sigma_0$. This constancy is guaranteed in dilute systems and further demonstrated in
Fig.~\ref{fig:stationary} for an example of concentrated charges.
The lower panel gives the periodic alternating electric field, $E^{\rm e3dtf}(z;[\rho^q])$ produced by the e3dtf potential for both fixed (black thick solid line,
$\pm\sigma_0$) and mobile (red dash line, $\pm\sigma_1$ and blue thin solid lines, $\pm\sigma_2$) surface charge densities.  See Eqs.~\ref{eq:e3dtffield}
and~\ref{eq:e_e3dtf}.
The area formed by any of the three electric fields and the $z$-axis (not shown) in any period of $L_z$ would be zero, indicating the compensating
boundary condition of Eq.~\ref{eq:compensating}. 
By matching the difference between the two SPMF potentials, $\nu_{\rm sp0}(z)$ and $\nu^{\rm e3dtf}_{\rm sp}(z)$ defined in Eqs.~\ref{eq:sp0}
and~\ref{eq:e3dtfsp}, the acting electric fields on both liquids are corrected to reestablish agreement with the macroscopic definitions of $\varepsilon_1$ and
$\varepsilon_2$, either of which could be $1$,  $>1$ but finite, or $\infty$ referring to the vacuum, a polar solvent, or an electrolyte respectively.}
\label{fig:model} \end{figure*}
On the other hand, a recent stimulating work by Zhang and Sprik (ZS) have shown that when the e3dtf method is used for charged insulator/electrolyte interfaces
(confined for the vacuum insulator and extended for the material insulator) in the $xy$ plane, the electric double layers (EDL) formed at the solid-electrolyte
interface are indeed not fully charge compensated because of the finite $L_z$ and hence the finite width of the insulator slab used\cite{Zhang_Sprik2016}. 
They have further applied the constant electric field methods\cite{Stengel2009} to derive a linear relation between the net EDL charge and the applied electric field
for given $L_z$. Their linear equation agrees excellently with the simulation data for the vacuum insulator (confined), however only qualitatively capture the
feature in the range of weak electric fields for the water insulator (extended).

The two previous work\cite{Pan_Hu2017,Zhang_Sprik2016} certainly call for a general treatment of the finite-size effect arising from $\nu^{\rm e3dtf}$ for extended
systems of charged interfaces. It is difficult to develop such a treatment because rigorous handlings of electrostatics for an extended system under the periodic
boundary condition (PBC) are not even conceptually available.
In this paper, we overcome this problem by conjecturing that a symmetry-preserving mean-field (SPMF) condition\cite{Hu2014spmf} must be satisfied by any accurate
method to reestablish agreement with macroscopic continuum electrostatics.
The e3dtf simulation of the response of coexisting dielectric fluids to charged walls and a constant electric field is then forced to match the SPMF condition by
intuitively adding both inbound and outbound corrections to the effective acting electric fields. The inbound correction reduces to the previous mean-field relation
for a confined system\cite{Pan_Hu2017} while the outbound correction successfully accounts for the nonlinear coupling between the two liquids.

In the next section, we derive mean-field equations correcting the response of the two dielectric fluids with relative permittivities of $\varepsilon_1$ and
$\varepsilon_2$ shown in Fig.~\ref{fig:model}. When the mobile charge densities of the two liquids are approximated as pairs of surface charge densities,
$\pm\sigma_1$ and $\pm\sigma_2$ separated by stationary distances $D_1$ and $D_2$ respectively, the mean-field equations reduce to simple analytic relations
among $\sigma_1$, $\sigma_2$, $\varepsilon_1$, $\varepsilon_2$, $D_1$, and $D_2$ under arbitrary conditions of the setup lengths $L_z$ and $D_0$, the fixed
surface charge density $\sigma_0$ and the external electric field $E_{\rm app}$.
No new simulations are carried out to validate the mean-field theory in this work. Instead, we reinterpret the simulation results in the literature to undoubtedly
demonstrate its predictions for three examples: (i) confined water, (ii) confined electrolyte and (iii) an extended system of coexisting water and
electrolyte. The well-justified mean-field approximations further suggest proper handlings of electrostatics for simulations of charged interfaces. Although we focus
on systems with perfect planar symmetry, we finally argue that the mean-field treatment of the finite-size effect is relevant for general biomolecular simulations.

\section{A mean-field theory correcting dielectric responses }
Macroscopic continuum electrostatics states that a constant electric field in the $z$ direction polarizes a dielectric fluid exposed to a solid surface in the $xy$
plane such that the mobile charges of the fluid reorganize to form a nonuniform charge density $\rho^q(z)$, that produces an internal polarization field
defined by integrating the Poisson's equation
\begin{equation} E_p (z;[\rho^q]) = \frac{1}{\varepsilon_0} \int^z dz^\prime\, \rho^q(z^\prime) , \label{eq:e_p}\end{equation}
where the integral takes a position at which $\rho^q$ vanishes (e.g. the location of the solid surface) as its lower bound. In the bulk region $z=z_b$ where
$\rho^q(z_b)$ vanishes again, $E_p(z_b;[\rho^q])$ partially cancels the external acting electric field. The factor by which the overall intrinsic electric field in
the bulk decreases relative to the acting electric field defines the relative permittivity (dielectric constant) of the material.

For the extended system of Fig.~\ref{fig:model} where the two surfaces confining the dielectric material $\varepsilon_1$ are in contact with the other dielectric
material $\varepsilon_2$, it is difficult to introduce a proper treatment of electrostatics such that both materials respond correctly to the apparent acting
electric fields. The SPMF approach claims that one may keep some short-ranged component of the Coulomb potential $\nu_0({\mathbf r})$ unchanged but average its
remaining slowly-varying long-ranged component in the $x$ and $y$ directions with preserved symmetry to achieve efficient and accurate simulations of structural,
thermodynamic and even dynamic properties\cite{Hu2014spmf}. Any accurate (acc) SPMF treatment, $\nu^{\rm acc}({\mathbf r})$ must therefore satisfy the SPMF
condition
\begin{equation}  \nu_{\rm sp0}(z) \equiv \left< \nu^{\rm acc}({\mathbf r})\right>_{\rm sp} = \frac{1}{A} \iint dxdy\, \nu_0({\mathbf r}) = 
 \frac{-|z|}{2\varepsilon_0 A}, \label{eq:sp0} \end{equation}
where the integral is taken over the infinite area for $\nu_0({\mathbf r})$ or $A=L_xL_y$ for $\nu^{\rm acc}({\mathbf r})$. $\left<\,\right>_{\rm sp}$  defines
the SPMF potential as the average over the degrees of freedom in the directions with preserved symmetry in general\cite{Hu2014spmf,Yi_Hu2017mf}. We now conjecture
that the SPMF condition serves as a guiding constrain in building any accurate treatment of electrostatics in a finite setup that is consistent with the above
macroscopic description of the dielectric response. 
In practice, $\nu^{\rm e3dtf}({\mathbf r})$ is commonly used for the supercell modeling of the extended system. The corresponding SPMF potential reads
\begin{equation} \nu^{\rm e3dtf}_{\rm sp}(z) \equiv \left< \nu^{\rm e3dtf}({\mathbf r}) \right>_{\rm sp} =\frac{\tau^{\rm 3D}}{4\pi\varepsilon_0} +
 \frac{1}{\varepsilon_0V}\sum_{s\neq 0} \frac{e^{isz}}{s^2}, \label{eq:e3dtfsp} \end{equation}
where $s=2\pi k_z/L_z$ with $k_z$ integers. $\nu_{\rm sp0}(z)$ and $\nu^{\rm e3dtf}_{\rm sp}(z)$ differ by a quadratic function for $z\in [-L_z,L_z]$
\begin{equation} \nu_{\rm sp1}(z) = \nu^{\rm e3dtf}_{\rm sp}(z) - \nu_{\rm sp0}(z) = z^2/(2\varepsilon_0 V)  +  C , \label{eq:sp1} \end{equation}
where the constant $C=\tau^{\rm 3D}/(4\pi\varepsilon_0) + L_z/(12\varepsilon_0A)$. They satisfy the Poisson's equation defined with two different charge densities
respectively, the unit surface charge density for the former
\begin{equation} - \nabla^2 \nu_{\rm sp0}(z) = \delta(z)/(\varepsilon_0 A), \end{equation}
and the periodic unit surface charge density with a compensating uniform background for the latter
\begin{equation} -\nabla^2 \nu^{\rm e3dtf}_{\rm sp}(z) = -\frac{1}{\varepsilon_0 V} + \frac{1}{\varepsilon_0 A} \sum_{n=-\infty}^\infty \delta(z - nL_z), 
\end{equation}
where $\delta(z)$ is the Dirac delta function. According to the superposition principle, the e3dtf electric field for any planar charge density $\rho^q(z)$ is given
by
\begin{equation} E^{\rm e3dtf}(z;[\rho^q]) \equiv - A\frac{d}{dz} \int_0^{L_z} dz^\prime\,\rho^q(z^\prime)\nu^{\rm e3dtf}_{\rm sp}(z-z^\prime),
\label{eq:e3dtffield} \end{equation}
where the interval of the integration, $[0,L_z]$ could always be an arbitrary period of $L_z$. $E^{\rm e3dtf}(z;[\rho^q])$ alternates within any period and satisfies
a compensating boundary condition
\begin{equation} \int_0^{L_z} E^{\rm e3dtf}(z;[\rho^q]) \equiv 0 \label{eq:compensating}, \end{equation}
which essentially reflects the periodicity of $\nu^{\rm e3dtf}_{\rm sp}(z)$. For the specific fixed charge density $\rho^q_f(z)$ represented by the two charged walls
separated by $D_0$ in Fig.~\ref{fig:model}, $E^{\rm e3dtf}(z;[\rho^q_f])$ gives {\it exactly} two constant electric fields for the regions inside and outside the
walls respectively
\begin{equation} E^{\rm e3dtf}(z_j;[\rho^q_f]) = \left\{  \begin{array}{lr}
-\sigma_0(L_z-D_0)/(\varepsilon_0L_z)  & j=1 \\
 \sigma_0D_0/(\varepsilon_0L_z)        & j=2  \end{array} \right.  \label{eq:e_e3dtf}\end{equation}
where $z_j$ ($j=1$ or $2$) denotes any $z$ position in the bulk of the material $j$. In contrast, the usual electric field for the two charged walls must be
$-\sigma_0/\varepsilon_0$ inside and $0$ outside.  $E^{\rm e3dtf}(z;[\rho^q_f])$ therefore involves a finite-size correction, $\sigma_0D_0/(\varepsilon_0L_z)$ for
both regions.

There exits an intuitive understanding of the finite-size effect caused by the nonuniform mobile charge densities of the two materials, $\rho^q_1(z)$ and
$\rho^q_2(z)$ as well.
Noting that $\nu_{\rm sp1}(z)$ is slowly varying relative to $\nu_{\rm sp0}(z)$ and short-ranged parts of other van der Waals interactions, the instantaneous
electric field from $\nu_{\rm sp1}(z)$ thus fluctuates weakly configuration by configuration in crowded environment\cite{Hu2014spmf} and can be safely approximated
by its equilibrium average
\begin{equation} E_c(z;[\rho^q]) \simeq - A \frac{d}{dz} \int dz^\prime\, \rho^q(z^\prime) \nu_{\rm sp1}(z-z^\prime), \end{equation}
where the range of the integration, if not stated, is always the whole continuous region defining $\rho^q(z)$. $z$ spans over the regions of the two
materials and $z-z^\prime \in [-L_z,L_z]$ necessarily. {\it The usual minimum image convention resulting in $|z-z^\prime|^{\rm min}\leqslant L_z/2$ is therefore not
applied here} ! As will be discussed later, this violation of PBC has an important consequence on the proper treatments of electrostatics. Anyway,
the simulation done with $\nu^{\rm e3dtf}({\mathbf r})$ is now effectively mapped into the simulation done artificially with $\nu_0({\mathbf r})$ subject to an
additional electric field $E_c(z;[\rho^q])$, that accounts for the average effect arising from their difference.
This static limit of the SPMF approximation resembles the local molecular field (LMF) approximation developed by Weeks and co-workers
\cite{Weeks2002,Rodgers_Weeks2008,Hu_Weeks2010lmf,Remsing_Weeks2016}. However, the present application differs from that of LMF 
theory in three aspects. (i) While LMF theory connects a full system with a long-ranged interaction to a simpler reference system with a short-ranged interaction,
both $\nu^{\rm e3dtf}({\mathbf r})$ and $\nu_0({\mathbf r})$ are long-ranged. (ii) When the relative distance increases, the usual long-ranged component in LMF
theory tends to $0$ but $\nu_{\rm sp1}(z)$ increases all the way up to $z=L_z$.
(iii) LMF theory often solves iteratively the spatially varying effective potential, this static approximation yields instead a constant electric field proportional
to the total dipole moment of the material
 \begin{equation} E_c(z;[\rho^q]) \equiv E_c([\rho^q]) = \frac{1}{\varepsilon_0L_z} \int dz^\prime\, z^\prime\rho^q(z^\prime), \label{eq:e_c}\end{equation}
when the electroneutrality condition, $\int dz\, \rho^q(z) = 0$ is applied. In total, each dielectric material $j$ ($j=1$ or $2$) responds effectively (eff) to a
sum of four constant electric fields 
\begin{equation} E^{\rm eff}_j = E_{\rm app} + E^{\rm e3dtf}(z_j;[\rho^q_f]) + E_c([\rho^q_1 + \rho^q_2]), \label{eq:e_eff}\end{equation}
where the last term must consist of an inbound correction (c), $E_c([\rho^q_j]) $ arising from the material $j$ itself and an outbound correction arising from the
other material. In addition, the outbound correction $E_c([\rho^q_{1/2}])$ reduces completely to $E^{\rm e3dtf}(z_{2/1};[\rho^q_{1/2}])$ and thus fully describes the
coupling between the two materials, which is consistent with the fact that the usual electric field due to $\nu_{\rm sp0}(z)$ always vanishes outside a neutral
material.
\begin{figure}[tbhp]
\centering
\includegraphics[width=0.8\linewidth]{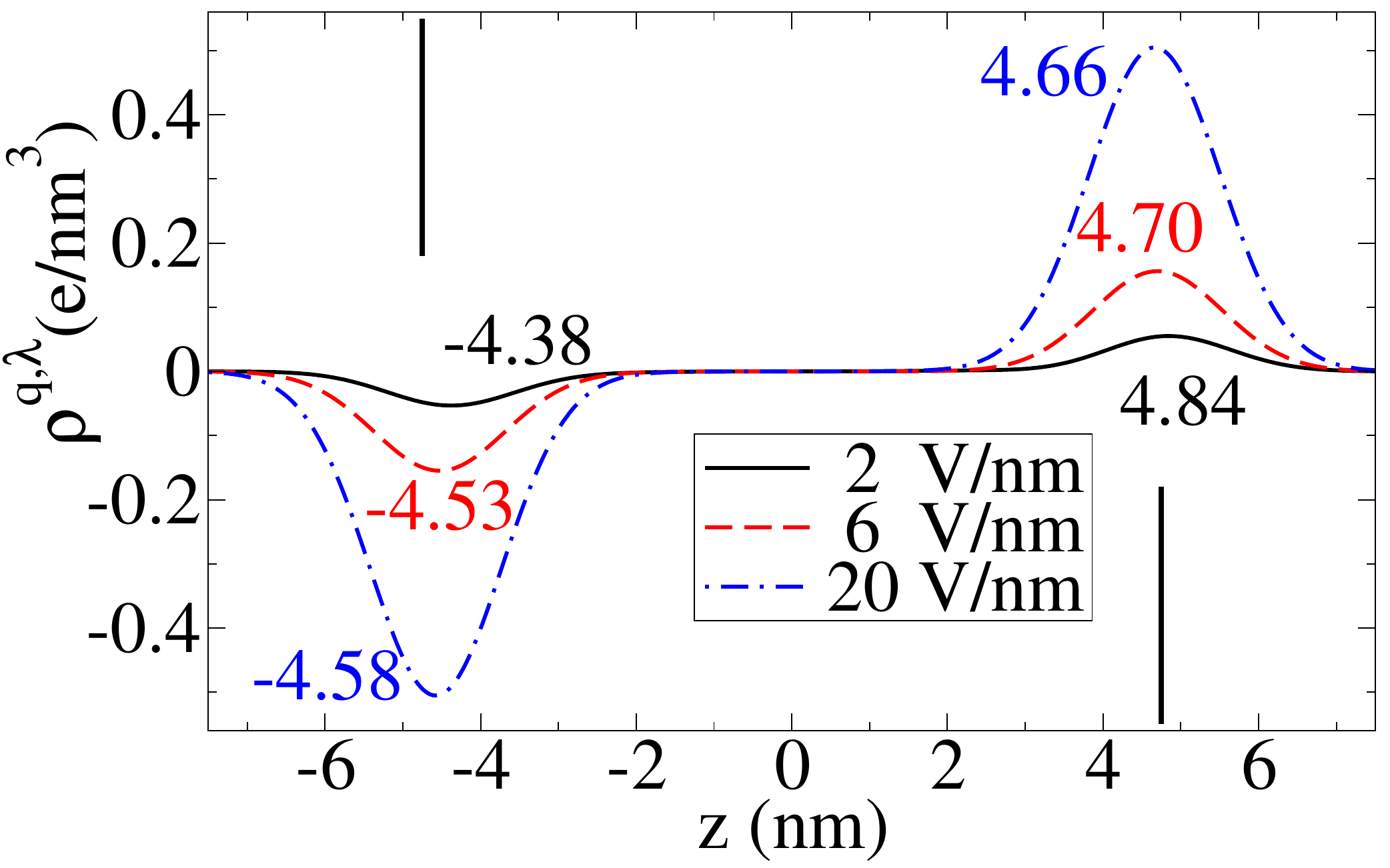}  
\caption{Smoothed charge densities $\rho^{q,\lambda}(z)$ of confined SPC/E water under the three acting electric fields of $2$ (solid line), $6$ (red dash line) and
$20$ (blue dot dash line) volts/nm simulated previously\cite{Hu2014spmf}. The $z$ positions of the peaks and troughs are
indicated by the nearby numbers. The two straight lines refer to the walls at $z=\pm 4.75$.
As a consequence of the very strong oscillations of the unsmoothed charge density under a weak electric field, one peak at $4.84\,{\rm nm}$ goes beyond the position
of the wall. $\rho^{q,\lambda}(z)$ shifts in the $z$ direction with an almost constant distance ($\simeq 9.22$ nm) between the two branches.
$\lambda=1.2\,{\rm nm}$ and similar observations are made for other values of $\lambda$. The walls are shifted outwards by $2.5\,{\rm nm}$ each for clarity.}
\label{fig:stationary} \end{figure}

The dielectric response of each material $\varepsilon_j$ corrected by the finite $L_z$ associated with the use of the e3dtf method follows
\begin{equation} E^{\rm eff}_j +  E_p (z_j;[\rho^q_j]) =  E^{\rm eff}_j / \varepsilon_j \quad{\rm for}\quad j=1\,\,{\rm or}\,\,2, \label{eq:cr} \end{equation}
which however is difficult to solve because $\rho^q_j(z)$ involved in Eqs.~\ref{eq:e_p} and~\ref{eq:e_c} oscillates strongly on the scale of intermolecular
distances. In fact, both equations can be re-expressed {\it exactly} as functionals of an auxiliary smoothed charge density, $\rho^{q,\lambda}(z)$:
$E_c([\rho^q]) = E_c([\rho^{q,\lambda}])$ and $E_p(z_b,[\rho^q])=E_p(z_b,[\rho^{q,\lambda}])$, where $\rho^{q,\lambda}(z)$ may be defined by convoluting 
$\rho^q(z)$ with a normalized Gaussian function with width $\lambda$ that exceeds a typical intermolecular distance of subnanometer
\begin{equation}  \rho^{q,\lambda}(z) = \int dz^\prime \, \rho^q(z^\prime) \frac{1}{\sqrt\pi\lambda} e^{-|z-z^\prime|^2/\lambda^2}. \end{equation}
Furthermore, when denoting the total surface charge density in the left side as $\sigma_j$, shown in Fig~\ref{fig:model}, and subsequently the surface charge dipole
moment as $-\sigma_j D_j$, $E_p(z_j;[\rho^q_j])=\sigma_j/\varepsilon_0$ and $E_c([\rho^q_j])=-\sigma_jD_j/(\varepsilon_0L_z)$. 
Given that the length of the material $j$ in the $z$ direction is sufficiently large ($>\lambda$) that the two interfaces of the same material are well separated by
the bulk, $D_j$ must stand for the distance between the two oppositely charged branches of any properly chosen $\rho^{q,\lambda}_j(z)$.
At the limit of dilute ions for which the linearized Poisson-Boltzmann theory applies, the charge density is proportional to $e^{-k_D z}$ with $k_D$ the inverse
Debye length\cite{Hansen_McDonald2013} and $D_j$ must be exactly independent of the acting electric field.
As an example of concentrated charges in a dense fluid, $\rho^{q,\lambda}(z)$ of confined SPC/E water\cite{Hu2014spmf} shown in Fig.~\ref{fig:stationary} suggests
that $D_j$ is still nearly a constant although the whole $\rho^{q,\lambda}(z)$ slips upon the change of the acting electric field. Any liquid-solid interface may
arguably be characterized by a stationary distance resulting from a balance between competing van der Waals and Coulomb forces.

Given the algebraic expressions of $E_p(z_j;[\rho^q_j])$ and $E_c([\rho^q_j])$, Eq.~\ref{eq:cr} turns into two coupled relations between the input parameters 
$E_{\rm app}$, $L_z$, $D_0$, and $\sigma_0$, and the unknown surface charge densities $\sigma_1$ and $\sigma_2$ for the given characteristics of $\varepsilon_1$,
$\varepsilon_2$, $D_1$ and $D_2$.
It is crucial that $D_j$ depends on neither $\sigma_0$ nor $E_{\rm app}$. Otherwise, Eq.~\ref{eq:cr} would not be analytically solvable.
Such simplification reflects essentially that the oscillations of $\rho^q_j(z)$ on the small intermolecular length scale hardly influence the integrated properties
on the much larger scale. 
\section{Applications: charged polar and electrolytic interfaces}
\begin{figure}[tbhp]
\centering
\includegraphics[width=0.8\linewidth]{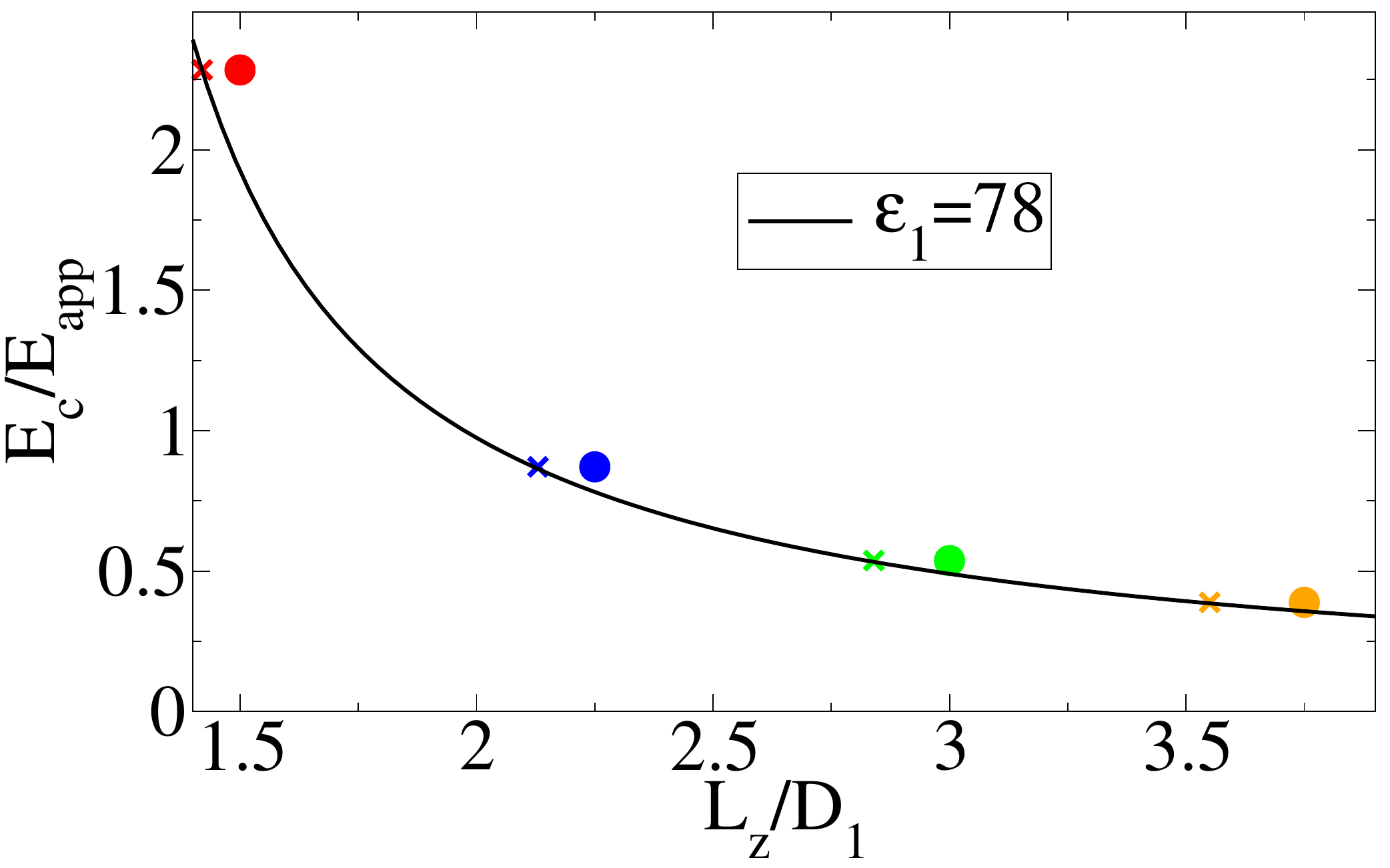}
\caption{Relative corrections to the applied electric field, $E_{\rm c}/E_{\rm app}$ plotted as a function of $L_z/D_1$. The solid line corresponds to
Eq.~\ref{eq:e_ce_p} with $\varepsilon_1=78$. Symbols indicate simulation results for $L_z=6$, $9$, $12$, and $15$ nm at $D_1=4.22\,{\rm nm}$ (crosses) or
$D_1=4\,{\rm nm}$ (filled circles taken from Fig.8 of ref\cite{Pan_Hu2017}).}  \label{fig:waterconfined} \end{figure}

We now examine the validity of the corrected dielectric response described by Eq.~\ref{eq:cr} for three charged interfaces of polar and/or electrolytic systems. 
In the first example of confined SPC/E water\cite{Berendsen_Straatsma1987} between the two planar walls\cite{Lee_Rossky1984} investigated previously by
us\cite{Pan_Hu2017}, $E_{\rm app}=2$ volts/nm, $\sigma_0=0$, $\varepsilon_1\simeq 78$ for SPC/E water under weak electric fields\cite{Pan_Hu2017}, and
$\varepsilon_2=1$ for the vacuum. Therefore, Eq.~\ref{eq:cr} only involves the inbound correction and thus reduces to Eq.(83) in ref.\cite{Pan_Hu2017}, which was
derived alternatively by analyzing the difference between the e3dtf method and the Ewald2D method.
However, the present derivation based on the SPMF constraint is more solid and generally applicable without addressing any specific treatment of electrostatics.
It has been verified that the simulated $\rho^q_1(z)$ using e3dtf with four different setups of $L_z$ indeed responds to the corresponding effective acting
electric field,  $E_c[\rho^q_1] + E_{\rm app}$. Moreover, without knowing $\rho^q_1(z)$, the inbound correction relative to the applied electric field,
$E_c([\rho^q_1])/E_{\rm app}$ is predicted to be
\begin{equation} E_c([\rho^q_1])/E_{\rm app} = (1-1/\varepsilon_1)/(L_z/D_1-1+1/\varepsilon_1), \label{eq:e_ce_p} \end{equation}
which is formally identical to Eq.(89) of ref.\cite{Pan_Hu2017} except that it becomes meaningful to assign an accurate value for $D_1$ now. Indeed, 
Fig.~\ref{fig:waterconfined} shows that this analytic relation agrees excellently with the simulation results when taking $D_1=4.22\,{\rm nm}$
rather than approximating $D_1\simeq 4\,{\rm nm}$ as in ref.~\cite{Pan_Hu2017}.

\begin{figure}[tbhp]
\centering \includegraphics[width=0.8\linewidth]{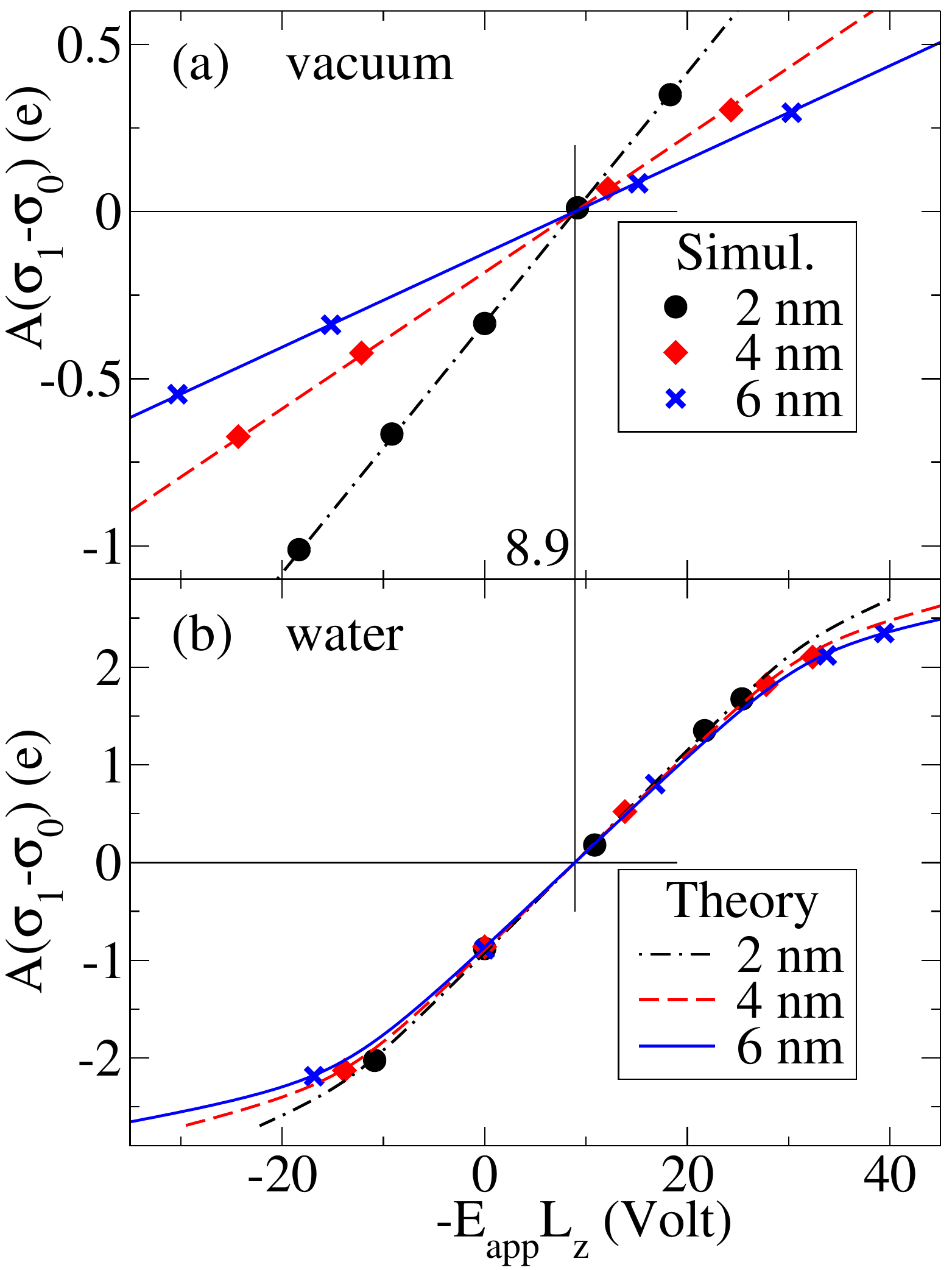}
\caption{The net EDL charge as a function of the external voltage for the vacuum (a) and water (b) insulators. Simulation data are taken from Fig. 6 and Fig. 8 of
ref~\cite{Zhang_Sprik2016} where $D_0=4.124\,{\rm nm}$, $L_x=L_y=1.275\,{\rm nm}$, $A\sigma_0=2 {\rm e}$, and $L_z$ equals $D_0$ plus the thickness of the
insulator slab: $2$ (circles), $4$ (diamonds), or $6$ (crosses) nm.
Analytic results are given by Eq.~\ref{eq:s_0s_1} at $D_0-D_1=0.4\,{\rm nm}$ and $L_z-D_0-D_2=0.445\,{\rm nm}$ with a reasonable expression, Eq.~\ref{eq:e_2} for the
field-dependent $\varepsilon_2$\cite{notedielectric}.
The straight line indicates the invariant zero net charge point $(8.9 {\rm V}, 0{\rm e})$ for any insulator. \label{fig:qvolt} } \end{figure}

The importance of determining precisely the value of $D_j$ is further demonstrated by applications of Eq.~\ref{eq:cr} to two other examples of insulator/electrolyte
interfaces. In the two EDL systems investigated previously by ZS, $\varepsilon_1=\infty$ for the electrolyte. Eliminating $\sigma_2$ from the coupled equations of
corrected dielectric response yields analytically
\begin{multline} (\sigma_1- \sigma_0) (L_z-D_1-D_2 + D_2/\varepsilon_2)= \\
 \varepsilon_0 (-E_{\rm app} L_z)  - \sigma_0 (D_0 - D_1) \label{eq:s_0s_1} ,\end{multline}
and $E_2^{\rm eff} = (\sigma_0 - \sigma_1)/\varepsilon_0$. For the vacuum insulator, $\varepsilon_2= 1$ as in the previous example. 
The slope of the linear relation between the net EDL charge, $(\sigma_1- \sigma_0)A$ and the applied voltage, $-E_{\rm app} L_z$ is predicted to be
$A\varepsilon_0/(L_z-D_1)$ with $D_1=3.724$ nm matching the unified $x$-intercept of $8.9$ volts found by ZS\cite{Zhang_Sprik2016}. This linear relation is confirmed
by the simulation results shown in Fig.~\ref{fig:qvolt}(a) for three setups of $L_z$. Alternatively, ZS have introduced a Stern-type model to interpret successfully
their findings for this vacuum insulator, which inspired us to develop the present mean-field approach to elucidate the underlying microscopic mechanism.

Indeed, such an attempt leads to even more satisfactory analysis of e3dtf simulation for any material insulator with $\varepsilon_2=\varepsilon_2(E_2^{\rm eff})$
depending on the acting electric field $E_2^{\rm eff}$ in general. For the SPC/E water insulator, one may use the expression derived by Booth for the field-dependent
relative permittivity\cite{Booth1951,notedielectric}
\begin{equation} \varepsilon_2 = n^2 + \frac{7\rho_b(n^2+2)\mu }{3\varepsilon_0 \sqrt{73} E_{\rm in}} L\left[\frac{\sqrt{73}\mu (n^2+2)
E_{\rm in}}{6 k_b T} \right], \label{eq:e_2} \end{equation}
where $E_{\rm in} = |E_2^{\rm eff}|/\varepsilon_2 = |\sigma_0-\sigma_1|/(\varepsilon_0\varepsilon_2)$ is the absolute value of the overall intrinsic electric field
exerted on the bulk.
$L(x)={\rm coth}(x) - 1/x$ is the Langevin function. $\mu=0.04894 {\rm e\cdot nm}$ and $\rho_b=33.46/{\rm nm}^3$ are the dipole moment and the number density at the
temperature $T=298 {\rm K}$ respectively. $k_b$ is the Boltzmann constant.
The refractive index is adjusted slightly, $n = 1.113$ such that the dielectric constant at a weak electric field $\simeq 78$. Because of the coupling between the
electrolyte and the material insulator, Eqs.~\ref{eq:s_0s_1} and~\ref{eq:e_2} suggest that the relation between the net EDL charge and the applied voltage is
nonlinear and symmetric with respect to the point of the $x$-intercept, $(8.9{\rm V},0 {\rm e})$, where the slope reaches the maximum,
$A\varepsilon_0/(L_z-D_1-D_2+D_2/78)$ and then decreases monotonically down to that of the vacuum insulator, $A\varepsilon_0/(L_z-D_1)$ as the absolute value of the
net EDL charge increases.
Fig.~\ref{fig:qvolt}(b) shows that this nonlinear relation agrees surprisingly well with the simulation results for the three setups of $L_z$ given that
$L_z-D_0-D_2=0.445\,{\rm nm}$ is a constant. 
More importantly, Eq.~\ref{eq:s_0s_1} predicts that the unified $x$-intercept representing the state of complete charge compensation, is invariant with respect to
solvent permittivities, which is undoubtedly confirmed by Fig.~\ref{fig:qvolt} for the two examples of insulator/electrolyte interfaces.

\section{Discussion: proper treatments for charged interfaces}
The present extended system may correspond to a very thin material $1$ ($D_0\sim$ nm) confined between the two charged walls, both exposed to the same macroscopic 
material $2$ from the other sides. In the $x$ and $y$ directions where the charges are uniformly distributed, the usual PBCs are safely applied. This section
discusses three boundary conditions: non-, full-, and semi-PBCs in the $z$ direction. 
\begin{figure}[tbhp]
\centering
\includegraphics[width=0.8\linewidth]{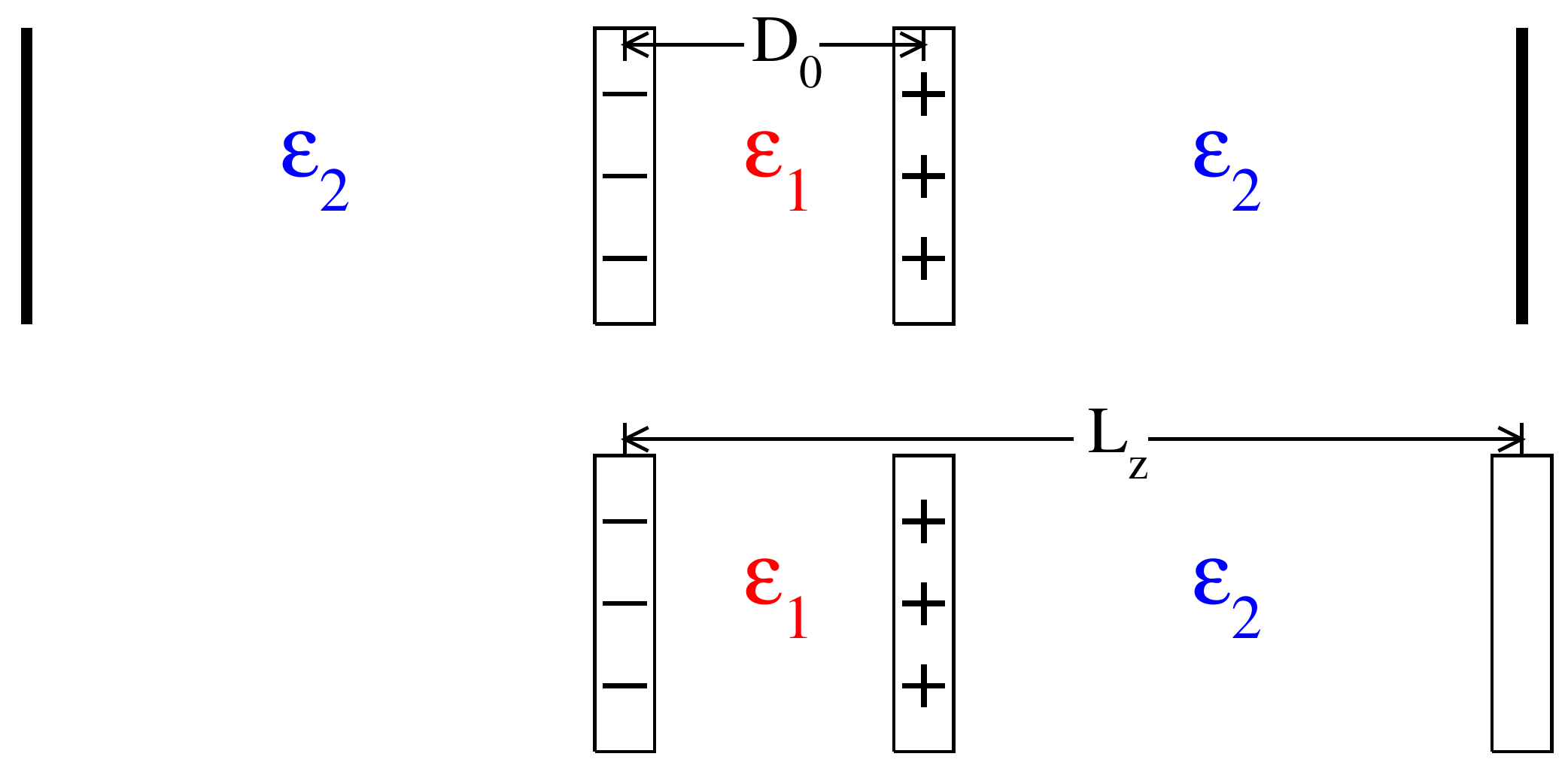}
\caption{Setups of non-PBC (upper) and semi-PBC (lower) in the $z$-direction in contrast with the full-PBC in Fig.~\ref{fig:model}. 
Under the semi-PBC, the minimum image convention due to PBC is only imposed on short-ranged interactions; 
charges appear instead continuously with no images when evaluating $z_{ij}$ of each pair for the long-ranged electrostatic interaction. \label{fig:pbcs}}\end{figure}

A natural way to mimic the realistic system is the non-PBC setup in Fig.~\ref{fig:pbcs}, under which two hard walls are introduced to turn the extended system into a
much larger confined system. Accurate algorithms for electrostatics in slab geometry include the Ewald2D sum methods\cite{Parry1975,Heyes_Clarke1977,
DeLeeuw_Perram1979,Pan_Hu2014,Pan_Hu2015}, the Ewald3D sum with the planar infinite boundary term methods\cite{Smith1981,Yeh_Berkowitz1999,Hu2014ib,Santos_Levin2016,
Yi_Hu2017pairwise}, and others\cite{Hautman_Klein1992,Boda_Henderson1998,Arnold_Holm2002,Hu2014spmf}, which all satisfy {\it exactly} the SPMF
condition in Eq.~\ref{eq:sp0}. 
Recent techniques developed for efficient simulations of bulk systems\cite{Wang_Fukuda2016,Vatamanu2018} can also be used for confined systems after adding an
extra $z$-dependent term to make the algorithms match the SPMF condition as previously done\cite{Hu2014spmf}.

An advantage of using the full-PBC in Fig.~\ref{fig:model} is to minimize the edge effect in a relatively smaller unit cell.
The present mean-field theory immediately suggests that the e3dtf method still works for simulating desired responses of both materials when two auxiliary constant
electric fields are applied on the materials to offset the unwanted components of $E_1^{\rm eff}$ and $E_2^{\rm eff}$ respectively. The auxiliary electric fields
depend on the setup and the intrinsic properties of $\varepsilon_1$, $\varepsilon_2$, $D_1$ and $D_2$, which can all be predetermined from trial simulations, as
evidenced by the perfect fittings in Figs~\ref{fig:waterconfined} and~\ref{fig:qvolt}.

Instead of applying statically predetermined auxiliary fields, it is convenient to offset the unwanted finite-size effect by adding instantaneously $-q_iq_j
z_{ij}^2/(2\epsilon_0V)$, the negative of Eq.~\ref{eq:sp1}, to each pair of charges. 
This instantaneous correction is formally identical to the planar infinite boundary term\cite{Smith1981,Hu2014ib} but should be interpreted in the setup of semi-PBC
in Fig.~\ref{fig:pbcs}, where $z_{ij}\in [-L_z,L_z]$ rather than $z_{ij}^{\rm min}\in [-L_z/2,L_z/2]$ due to the PBC.
Because $\nu^{\rm e3dtf}(z_{ij})=\nu^{\rm e3dtf}(z_{ij}^{\rm min})$ remains unchanged, the semi-PBC corresponds to applying PBC only for short-ranged interactions and
therefore interprets appropriately the SPMF condition. Besides, it is more accurate in principle than the above full-PBC scheme because the involved SPMF approximation
is always superior to its static limit\cite{Hu2014spmf,Yi_Hu2017mf}. 
The aforementioned other techniques for confined systems are applicable as well provided that $z_{ij}$ involved in the techniques is correctly evaluated under the
same semi-PBC.
To our knowledge, the semi-PBC scheme has not been implemented for any extended system before but has both advantages of using a smaller unit
cell and carrying out no trial simulations.

\section{Conclusions and Outlook}
We have demonstrated that the SPMF condition serves as a necessary constrain for any accurate handling of electrostatics in systems of charged
interfaces. The mean-field equation accounting for the difference between the usual e3dtf method and the SPMF condition is useful for providing transparent analysis
of the associated finite-size effect. 
Proper analysis of e3dtf simulations of any extended polar or charged system is expected to validate undoubtedly the analytic predictions from the mean-field
equation.
In addition, the mean-field theory suggests an efficient and accurate semi-PBC scheme that reconciles directly the simulated charge reorganization with the
macroscopic dielectric response. We expect this approach to be of significant importance to the study of large interfaces of materials and biological membranes with
their surface corrugations on a scale where the long-ranged electrostatics varies sufficiently slowly.

Although the present theory is based on classical point charge models, we are optimistic that the mean-field idea can be extended to both quantum mechanical
calculations\cite{Cheng_Sprik2014} and sophisticated models involving electric multipole potentials\cite{Li2016}.  
As such, we view the present work as a demonstration of the ability of the SPMF approach to treat accurately
and to understand analytically electrostatics in various complex molecular interfaces.

\section{Acknowledgement}
This work was supported by the NSFC Grant (No.21522304) and the Program for JLU Science and Technology Innovative Research Team (JLUSTIRT).


\begin{thebibliography}{40}
\bibitem []{notedielectric}%
  \BibitemOpen
  \href@noop {} \bibinfo {note} {In the present case,
  $\varepsilon_2(E_2^{\rm eff})$ is still quite large at the moderately strong
  acting electric fields. $1/\varepsilon_2<0.1$, thus contributes little to
  Eq.~\ref{eq:s_0s_1} such that $\sigma_1-\sigma_0$ becomes insensitive to a
  detailed expression of the field-dependent dielectric constant, like
  Eq.~\ref{eq:e_2}, which was derived originally for realistic water but
  employed here for a classic SPC/E model of water. It is otherwise necessary
  to use an accurate expression of $\varepsilon_2(E_2^{\rm eff})$ for the
  modeled fluid when $1/\varepsilon_2$ approaches $1$ under very strong acting
  electric fields.}\BibitemShut {Stop}%
\end{thebibliography}

%
\end{document}